\documentstyle[12pt]{article}
\input psfig.sty
\def\reference{\parskip 0pt\par\noindent\hangindent 0.5 truecm}
\def\kms{km ${\rm s}^{-1}$}
\def\R  {{\cal R}}                     
\def\ebar {\bar{\eta}}               
\def\msun{{\rm M_\odot}}             
\def\kms {\,\rm{km\,s}^{-1}}     
\def\ergs {\,\rm{erg\,s}^{-1}}   
\def\spose#1{\hbox to 0pt{#1\hss}}
\def\lta{\mathrel{\spose{\lower 3pt\hbox{$\mathchar"218$}}
     \raise 2.0pt\hbox{$\mathchar"13C$}}}
\def\gta{\mathrel{\spose{\lower 3pt\hbox{$\mathchar"218$}}
     \raise 2.0pt\hbox{$\mathchar"13E$}}}
\begin{document}
%
%
\title{Luminous ``Dark'' Halos}
%


\author{Mark A. Walker
} 

\date{}
\maketitle

{\center
Research Centre for Theoretical Astrophysics,\\School of Physics A28,
University of Sydney, NSW 2006\\m.walker@physics.usyd.edu.au\\[3mm]
}

%
\begin{abstract}
Several lines of evidence suggest that cold, dense gas clouds
make a substantial contribution to the total mass of dark halos.
If so then physical collisions between clouds must occur; these
cause strong, radiative shocks to propagate through the cold gas,
with the startling implication that all ``dark'' halos must be luminous.
The expected luminosity is a strong function of halo velocity dispersion,
and should contribute a significant fraction of the observed X-ray
emission from clusters of galaxies, if dark halos are predominantly
made of cold gas.

Existing data do not exclude this possibility; indeed two
particular expectations of the luminous-halo model are borne
out in the X-ray data, and thus give support to the cold-cloud
dark matter model. First we find a luminosity-temperature
correlation of the form $L\propto T^{11/4}$, as seen in recent
analyses of cluster samples. Secondly the anticipated spectra have
substantially more power at low energies than isothermal bremsstrahlung
spectra, and might account for the observed ``excess'' EUV emission seen
from some clusters. The successes of the luminous-halo model are
particularly remarkable because the theory has no free parameters or
ad hoc elements. The model can be tested by
the X-ray satellite Chandra, which should resolve the Virgo cluster
into $10^4$ point-like, transient X-ray sources. Non-detection of
any such sources by Chandra can constrain the contribution of
cold gas clouds to be $\lta1$\% of the total matter density in
the Universe, assuming Virgo to be representative.
\end{abstract}


\bigskip

\section{Introduction}
Several years ago it was proposed that cold gas could make up
a significant fraction of the dark matter in spiral galaxies
(Pfenniger, Combes \& Martinet 1994). This particular
proposal advocated massive, fractal gas clouds distributed in
a thin disk, but subsequent authors have contemplated spherical
clouds in the (dynamically more conventional) context of a
quasi-spherical halo (Henriksen \& Widrow 1995; de~Paolis et~al
1995; Gerhard \& Silk 1996; Walker \& Wardle 1998). Walker \&
Wardle's (1998) model for Extreme Scattering Events (radio wave
lensing events) requires that essentially {\it all\/} of the
Galactic dark matter be in the form of cold, dense gas clouds.
Data on Galactic emissions -- notably in the $\gamma$-ray region
(de~Paolis et al 1995; Kalberla, Shchekinov \& Dettmar 1999)
-- and LMC microlensing properties (Draine 1998; Rafikov \& Draine
2000) do not exclude this possibility. A natural question, then,
is to ask whether all of the dark matter in all halos (galaxies and
clusters of galaxies) might assume the form of cold gas. However,
this extension precipitates a conflict with established ideas,
because clusters of galaxies are generally assumed to be
so large that they are representative samples of the Universe,
and well-known arguments favour a predominantly non-baryonic
Universe (e.g. Peebles 1993). One possible resolution of this conflict
has been considered by Walker \& Wardle (1999), who emphasised the
loopholes in the case for non-baryonic dark matter. Even in the absence
of a resolution, though, it is useful to contemplate purely
baryonic models of dark halos, in order to clarify the strengths
and weaknesses of these models; that is the spirit of this paper.

Three discoveries in the
last two years have promoted the basic idea of dark matter in the form
of cold, dense gas clouds. First Walker \& Wardle (1998)
were able to explain the enigmatic ``Extreme Scattering
Events'' (Fiedler et al 1987) as radio-wave lensing events
caused by the photoionised surfaces of cold clouds. This model requires
``lens'' radii of order 2~AU, individual masses in the planetary range,
and a total mass which dominates the mass of the Galaxy. Secondly,
Dixon et al (1998) found that the $\gamma$-ray background contains
a substantial component attributable to the Galactic halo; given that
the diffuse gas in the Galactic plane is the principal feature of the
$\gamma$-ray sky (e.g. Bloemen 1989), this is prima facie evidence for
unseen gas in the Galactic halo (de~Paolis et al 1995). Thirdly,
Walker (1999: W99) showed that the cold-cloud model predicts a
relation between visible mass and halo velocity dispersion,
$M_{vis}\propto\sigma^{7/2}$, which agrees extraordinarily well with
data on spiral galaxies. Indeed this result appears to underlie
the Tully-Fisher relation, with the latter following when most of
the visible mass is in stellar form. In the model of W99 these results
arise from consideration of the collisions which must occur between
clouds (Gerhard \& Silk 1996); such collisions destroy the colliding
pair, and in this picture the visible content of any halo increases
with time as dark matter is converted to visible forms.

The success of this simple picture of (visible) galaxy assembly
encourages further investigation into the physics of cloud-cloud
collisions. One of the most basic features of the collision process
is that it involves strong shocks in the cold gas. By virtue of
the high particle densities ($\sim10^{12}\,{\rm cm^{-3}}$) within
the clouds, the post-shock radiative cooling time-scales are very
short and the shocks are radiative. This means that the bulk of the
kinetic energy dissipated during a collision goes into radiation,
implying a minimum level of emission from ``dark'' halos. This
is a key prediction which must be squared with the data: is the
model in conflict with observations? Here we investigate that question
and answer in the negative.

The model does lead to a highly unconventional picture of the
origin of X-ray emission in clusters of galaxies, and some readers
will regard this as intrinsically problematic because they believe
the phenomenon to be well understood. However, in both the proposed
picture and the conventional one, X-ray emission arises from a
two-body collision process acting in an isothermal gas in virial
equilibrium in the cluster potential; consequently the two theories
are degenerate in many respects, as will be evident in \S3. One
aspect of the new theory is, however, so strikingly different from
the accepted interpretation that there will be no difficulty in
deciding between the two, once new data are acquired. Because of
the simplicity with which this issue will be decided, no attempt
is made here to model anything other than the fundamental properties
of the X-ray emission. These properties are consistent with
existing data and, bearing in mind that the new theory has no
free parameters, it seems appropriate to accept the model as a
bona fide alternative for the time being, pending the outcome
of the test described in \S4.

We follow W99 in modelling dark halos as isothermal spheres which
are entirely composed of dense gas clouds;
we adopt a Maxwellian distribution function, and for simplicity we
assume that all clouds have the same mass and radius. The
expected properties of the emission are presented in the next
section. Because the predicted luminosity is a steep function of
halo velocity dispersion, we then (\S3) focus on the application
to clusters of galaxies, where intense X-radiation is expected.
Implications of the theory and ways in which it can be tested are
given in \S4.

\section{Basic properties of the emission}
In order to calculate the emissivity, $\varepsilon$, of a halo we
need to estimate the rate at which kinetic energy is dissipated in
collisions between clouds. Essentially all collisions are highly
supersonic, so we can apply conservation of momentum to each element
of area, $\Delta A$, of the two colliding clouds, with local surface
density $\Sigma_1,\Sigma_2$ (these surface densities being measured
parallel to the relative velocity vector). If each cloud has speed
$u$, in the frame of the centre-of-mass, before the collision, then
for a fully inelastic collision the final speed is just
$u\,|\Sigma_1-\Sigma_2|/(\Sigma_1+\Sigma_2)$. The change in kinetic
energy in each elemental area is thus $2\Delta A\,\Sigma_1\Sigma_2
u^2/(\Sigma_1+\Sigma_2)$. If we define $\eta$ to be the total kinetic
energy dissipated as a fraction of the total initial kinetic energy
(in the centre-of-mass frame), then we have
$$
\eta={{1}\over{M}}\int{\rm d}A\;{{2\,\Sigma_1\Sigma_2}\over
{\Sigma_1+\Sigma_2}}\,,\hfill\eqno(1)
$$
where $M$ is mass of a cloud. Evidently $\eta=\eta(b)$ is a function
of impact parameter, $b$, for the collision, and depends on the
density profile within the cloud. We shall be primarily concerned with
the average value $\ebar$:
$$
\ebar = {{1}\over{2r^2}}\int_0^{2r}{\rm d}b\,b\,\eta(b)\,,\hfill\eqno(2)
$$
where $r$ is the cloud radius. We have evaluated $\ebar$ for
polytropic cloud models of indices $n=1.5,\; 3,\; 4$, with the results
$\ebar=0.136,\; 5.37\times10^{-2},\;1.66\times10^{-2}$ respectively.

A firm model for the density profile of the putative dark clouds has
not yet been constructed. Wardle \& Walker (1999) suggest that solid
molecular hydrogen plays a key r\^ole in their thermal regulation,
in which case most of the radiative losses are likely to occur
from a thin surface layer because the precipitation/sublimation
balance is very temperature sensitive. Beneath this radiative
layer the dominant cooling is expected to come from spectral lines
which are very optically thick, and we anticipate that these regions
are thus unstable to convection (Clarke \& Pringle 1997). We
therefore adopt a polytropic model with $n=1.5$ as an approximation
to the likely cloud density structure, leading to $\ebar\simeq0.136$.

In deriving the rate of collisions between clouds, $\R$, we assume
that the cloud population follows a Maxwellian velocity distribution
with total density $\rho$, leading to (W99)
$$
\R = {{16}\over{{\sqrt{\pi}}}}{{\rho^2\sigma}\over{M\Sigma}}\,,\hfill\eqno(3)
$$
($\Sigma$ is henceforth the average surface density of a cloud), with a mean
kinetic energy of $2\ebar M\sigma^2$ dissipated per collision. (Note
that the numerical coefficient in eq. 3 differs slightly from W99's
eq. 1, because we have specified a Maxwellian distribution
function. Similar, slight differences will be evident when
comparing some of our subsequent results with those of W99.) It is
now trivial to determine the local emissivity of the halo:
$\varepsilon=2\ebar M\sigma^2\,\R$. We employ W99's eq. 3
for the halo density distribution -- implicitly assuming that the dark
halo is entirely made up of cold clouds -- whence the intensity
$$
I = {{1}\over{4\pi}}\int{\rm d}s\,\varepsilon=
{{\ebar}\over{64}}\left[{{\sigma^5\Sigma}\over{G\,t^3\sqrt{\pi}}}
\right]^{1/2}{{1}\over{(1+x^2)^{3/2}}}\,,\hfill\eqno(4)
$$
where $x$ is the projected distance of the line-of-sight
from the centre of the halo, in units of the core radius,
$r_c$, and $r_c^2 = 16\sigma^3t/\pi^{3/2}G\Sigma$. Here $t$ is the
time which has elapsed since the halo virialised; we adopt
$t\simeq10$~Gyr, corresponding to halos which virialised
at redshifts $z\gta1$. The total luminosity
can be found simply by integrating eq. 4, leading to
$$
L=8\pi^2r_c^2\int_0^\infty{\rm d}x\,xI(x)=
2\ebar\left[{{\sigma^{11}\sqrt{\pi}}\over{G^3\Sigma\,t}}
\right]^{1/2}\,.\hfill\eqno(5)
$$
This result may also be written as $L=\ebar\dot M_{vis}\sigma^2$,
where $M_{vis}=\pi\sigma^2r_c/G$, emphasising the connection
with the pseudo-Tully-Fisher relation derived by W99. The average
column density of the individual clouds can be measured by fitting
the theoretical relation $M_{vis}(\sigma)$ to data for spiral
galaxies (W99); for our Maxwellian distribution function this
yields $\Sigma=134\;{\rm g\,cm^{-2}}$ for $t=10$~Gyr. We can
now evaluate eq. 5 numerically:
$$
L\simeq3.2\times10^{44}\sigma_3^{11/2}\qquad\ergs\,,\hfill\eqno(6)
$$
where $\sigma=10^3\,\sigma_3\;\kms$; this implies very luminous
halos for clusters of galaxies ($\sigma_3\sim1$).

The equality $L=\ebar\dot M_{vis}\sigma^2$ is important because
it demonstrates a close tie between the predicted luminosity and
the observed Tully-Fisher relation. That is, if W99's theory is
a correct explanation for the Tully-Fisher relation, then a
result very similar to equation (6) must follow for the
bolometric luminosity of the halo; this is independent
of the value of $\Sigma$, or the number density and spatial
distribution of the clouds. 

\begin{figure}
\begin{center}
\centerline{\psfig{file=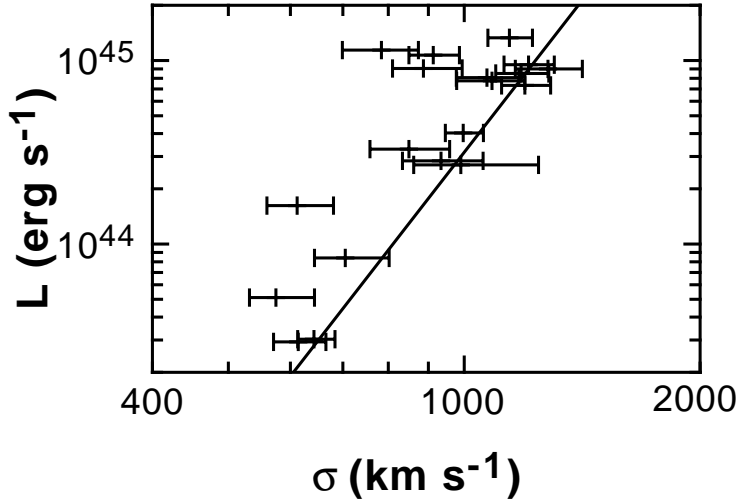,height=7cm}}
\caption{Theoretical $L(\sigma)$ relation from eq. 6 (solid line)
	together with data for 19 clusters. Luminosities are from Arnaud
	\& Evrard (1999) [Virgo, A262, A1656, A2634, A1060, A3558], and
	Markevitch (1998) [A85, A119, A399, A401, A754, A1795, A2256,
	A3266, A3391, A3395, A3571, A3667, MKW3S]; a Hubble constant of
	$H_0=75\;{\rm km\,s^{-1}\,Mpc^{-1}}$ is assumed. Velocity
	dispersions are from Girardi et al (1996). Errors in
	luminosity are expected to be small in comparison with those
	in $\sigma$.}
\end{center}
\end{figure}

What about the spectrum of the radiation? For the present it suffices
to note two general points. First, the radiation is thermal; and
secondly, a fiducial temperature for the radiation is that of
the shocked gas. This temperature can be estimated from the jump
conditions for a strong shock: $kT_s=(3/16)\mu u^2$, where $\mu$
is the mean molecular mass. There is no unique value of the
shock speed, $u$, but $\langle u^2\rangle=2\sigma^2$ so
$kT_s\sim2\,\sigma_3^2$~keV, and this gives us a crude measure
of the typical photon energy. In this way we see that the halos
of dwarf galaxies ($\sigma\lta50\;\kms$) should emit mostly in
the optical and near-IR; this radiation is observable
in principle, but we note the low luminosities implied by
eq. 6 ($L\lta2\times10^{37}\;\ergs$).
Normal and giant-galaxy halos should emit mainly far- and
extreme-UV, which is not ordinarily observable because of the
large opacity of the Galactic interstellar medium in these bands.
The halos of clusters of galaxies should emit X-radiation which
is both observable and at a level which is easy to detect.
Consequently we expect that clusters
offer the best prospects for testing our theory, and we now
focus our attention on these systems.

\section{X-ray emission from clusters of galaxies}
The first question we must address is whether the predicted
luminosity is consistent with the data for clusters. Conventionally
the observations are interpreted in terms of two components: one due
to hot gas spread throughout the cluster, and another due to a central
``cooling flow'' (e.g. Fabian 1994).
Cooling flows introduce a large scatter in the observed
luminosity-temperature correlation (Fabian et al 1994). Our model
involves X-ray emission arising from cloud collisions
throughout the cluster, not just the
central regions, and must be compared with the cluster-wide component;
it is this component which is meant henceforth when we refer to the data.

The systematic trend of luminosity with X-ray spectral temperature,
$L(T_X)$, has been the subject of several recent studies (Markevitch
1998; Arnaud \& Evrard 1999; Reichart, Castander \& Nichol 1999),
with very similar results: $L_{bol}\propto T_X^{2.80\pm0.15}$
(Reichart et al 1999). In our model all temperatures scale with
$\sigma^2$, so eq. 5 implies a close parallel with the data:
$L_{bol}\propto T_X^{11/4}$. We note also the study of
Wu, Xue \& Fang (1999) which, although it did not exclude the
cooling flow contribution to $L_{bol}$, employed such a large
sample of clusters that the deduced correlation was nevertheless
very precisely determined: $L_{bol}\propto T_X^{2.72\pm0.05}$,
in agreement with the theory we have presented.

For clusters which have detailed
optical spectroscopy in addition to the X-ray data, we can assess
the dependence of $L$ directly on $\sigma$, as measured from cluster
galaxy velocity dispersions. Contamination by field galaxies, small
sample sizes, sub-clustering and anisotropic velocity distributions all
mean that measuring $\sigma$ is not easy. Girardi et al (1996) have
made careful estimates of $\sigma$ in 38 rich clusters; their
sample has 13 and 6 clusters in common with the samples of
Markevitch (1998) and Arnaud \& Evrard (1999), respectively. Taking
bolometric luminosities (for $H_0=75\;\kms{\rm\,Mpc^{-1}}$)
from the latter data sets, and the velocity dispersions determined
by Girardi et al (1996), we arrive at the points shown in figure 1.
Also shown is the theoretical prediction given by eq. 6,
from which we see that the data are all consistent with or
in excess of the prediction. In the context of our model,
a measured luminosity which is significantly in excess of the
prediction must be interpreted in one of two ways: as emission
from diffuse hot gas, spread throughout the cluster (see \S4),
or as a halo which virialised relatively recently (cf. eq. 5),
e.g. in a cluster merger event.

Is the spatial distribution of emission within
clusters consistent with our theory? Our model has a mean intensity
profile (eq. 4) which is identical to the standard model profile
(e.g. Sarazin 1988) $(1+x^2)^{-3\beta+1/2}$ with $\beta=2/3$; this
is an adequate approximation for many clusters (Jones \&
Forman 1984). Exact agreement
should not be expected because an isothermal sphere is only a
crude approximation to the likely dark halo density distribution.
A more realistic model distribution might include many smaller halos,
perhaps associated with individual cluster galaxies (cf. Moore
et al 1999) and these would give local enhancements in the mean
X-ray intensity, with modest attendant spectral changes.
An important aspect of the
present model is that it predicts a graininess in the intensity
profile, at any instant, because the total cluster emission is
contributed by a large number of discrete sources. This feature
is fundamental to the theory and admits an unequivocal
test with high resolution imaging data, as discussed in \S4.

Attempting to predict the spectra resulting from cloud collisions
is a formidable task. Consider first that the unshocked gas (density
$\sim10^{12}\;{\rm cm^{-3}}$, and temperature of several Kelvin) is,
initially, entirely opaque to X-rays as a result of the bound-free
opacities of hydrogen and helium. Because of the high temperature
of the shocked gas, the resultant photons ionise the upstream
material (cf. Shull \& McKee 1979), thus erasing the principal source
of opacity.  For collisions occurring in clusters the mean energy
dissipated per unit cloud mass is so large, roughly $100\,\sigma_3^2$
times the total chemical binding energy of the cold gas, that
we expect the ionisation fronts to break out of the clouds very
quickly. Thereafter the primary opacity presented by the unshocked
gas is due to electron scattering. Each X-ray photon is expected to
scatter hundreds of times before escaping, with a few eV exchanged
between electron and photon on each scattering. Thus, although the
thermal coupling is loose, in total there is a significant exchange
of energy between the escaping photons and the upstream gas. Add
to this the complex, time-dependent geometry associated with shocks
in a pair of colliding clouds, and we see that it will not be easy
to arrive at reliable quantitative predictions of the spectra even
for single collisions. The observed spectrum is, of course, a
sum of the spectra of a large number of collisions, with a spread
in collision speed; but this aspect of the calculation is more
straightforward, as the cloud kinetics are likely to be reasonably
well approximated by a Maxwellian distribution function.

It is beyond the scope of this paper to attempt a prediction of the
spectrum, instead we confine our attention to a single qualitative
point: the observed spectra should exhibit a strong low-energy component.
One can easily see that such a component should be present because
the post-shock gas cools as it flows downstream, and emission from
this gas will be predominantly in the soft X-ray and EUV bands. To
illustrate this point we have calculated an idealised spectrum
which neglects radiative transfer through the upstream gas. This
calculation assumes: a Maxwellian cloud distribution function;
the strong shock limit (cold upstream gas); pure bremsstrahlung
emission; and the optically thin limit. The resulting spectral
form is given by
$$
{{\nu L_\nu}\over{L}}={{\omega}\over{\gamma-1}}\int_{\xi_1}^\infty
 {{{\rm d}\xi}\over{\xi}}\left[\gamma-{{1}\over{\xi-1}}\right]
  {\cal S\/}(p)\,,\hfill\eqno(7)
$$
where $p\equiv \omega\xi^2/(\xi-1)$, $\omega\equiv h\nu/\mu\sigma^2$,
$\xi_1=(\gamma+1)/(\gamma-1)$, $\gamma=5/3$ and
$$
{\cal S\/}(p)=\int_0^\infty {\rm d}q\;q\,\exp(-q-p/q)\,.\hfill\eqno(8)
$$
The spectrum of eq. 7 is shown in figure 2, along with a
bremsstrahlung spectrum from isothermal gas with
$kT=\mu\sigma^2$, representing the conventional theory of
cluster X-ray emission. Relative to the standard theory it can
be seen that this calculation predicts a much broader spectrum
which peaks at lower energies, with a much larger fraction of
the power emerging at $\omega\ll1$.

\begin{figure}
\begin{center}
\centerline{\psfig{file=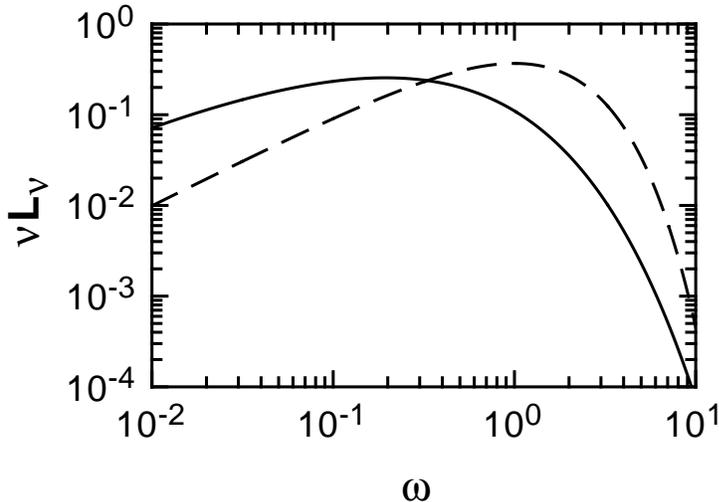,height=7cm}}
\caption{Illustrative theoretical spectrum due to cloud-cloud
    collisions (solid line, corresponding to eq. 7), and for comparison an
	exponential spectrum (dashed line, corresponding to free-free emission
	from an isothermal gas). Both spectra have unit bolometric luminosity.
	Notice that the spectrum for collisions has a great deal of power
	emerging at $\omega\ll1$, i.e. in the EUV band.}
\end{center}
\end{figure}

We emphasise that this calculation is only intended to be illustrative;
the assumptions employed are {\it not\/} good approximations to
the actual physical conditions, and the computed spectrum is therefore
not quantitatively correct. However, the qualitative
point that a high EUV luminosity is expected, relative to the
X-ray luminosity, should be model independent. This result is of
particular current interest as it has recently become apparent that
some clusters have EUV luminosities which are {\it much\/}
higher than expected on the basis of an isothermal bremsstrahlung
model for the X-ray emission (Mittaz, Lieu \& Lockman 1998; Lieu,
Bonamente \& Mittaz 1999). It is not currently known whether this
difficulty extends to all clusters. Various models have been proposed
specifically to account for these EUV data (e.g. Sarazin \& Lieu
1998), but dark matter in the form of cold clouds may be able to
explain this emission without the need for such ad hoc introductions.

The implication of a relatively large luminosity at low energies
raises the question of whether the proposed model is consistent
with the known X-ray spectra of clusters, which for the most part
are well described by optically thin emission from a single-temperature
hot gas. In particular, the model demands a somewhat surprising
coincidence whereby a complex amalgam of physics leads to apparently simple
X-ray spectra. Unfortunately this issue is difficult to address because
it requires a detailed computation of the spectral shape. By contrast
one can confidently assert a high EUV luminosity, because even a modest
difference in spectral index, over a large range in photon energy,
will manifest itself as a significant difference in flux. One should,
therefore, also expect real clusters to show significant departures
from the conventional model at very high X-ray energies, although it
is not clear whether an ``excess'' or a ``deficit'' is to be expected
at these energies. (One would, for example, need to know the precise
form of the the dark matter distribution function in order to
decide this question.) Observationally, studies at very high X-ray
energies are difficult because of the paucity of photons, but in
at least some cases, e.g. the cluster A2199 (Kaastra et al 1999),
there is evidence of an excess relative to the conventional model.
We emphasise that for the model we have presented, the issue
of the detailed spectral shape is not a critical one at present,
because a powerful test of the theory is
available via the predicted spatial distribution (see \S4).

\section{Discussion}

\begin{figure}
\begin{center}
\centerline{\psfig{file=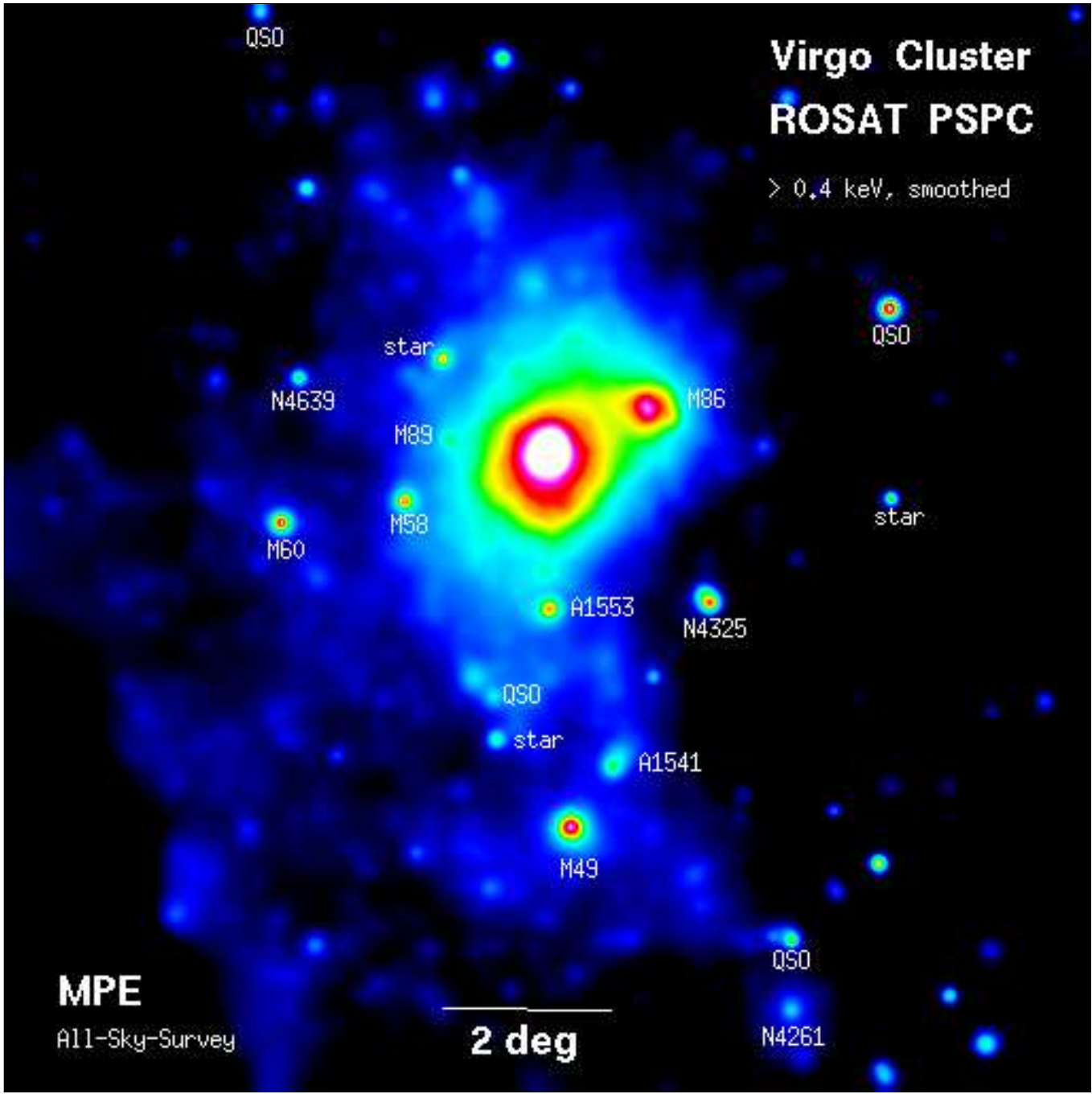,height=18cm}}
\caption{ROSAT All Sky Survey image of the Virgo cluster
of galaxies. These X-ray data clearly manifest a high
level of compact substructure near the periphery of the
cluster, consistent with the presence of a large number
of point-like sources. This image is a data product of
the ROSAT Mission (http://www.xray.mpe.mpg.de/), courtesy
Max-Planck-Institut f\"ur extraterrestrische Physik.}
\end{center}
\end{figure}

It is important to recognise that the theory presented in
\S\S2,3 is not incompatible with hot, diffuse
gas contributing to the observed X-ray emission, rather
the opposite in fact. W99 computed the total visible mass
which should accumulate within a halo of given velocity
dispersion, as a consequence of cloud-cloud collisions
(all of which disrupt the cold clouds):
$M_{vis}=7.4\times10^{13}\,\sigma_3^{7/2}\;\msun$ after
an interval of 10~Gyr. W99 gave no predictions as to what
form this visible material should take (e.g. stars vs. diffuse
gas). The material released by collisions is initially just
gas in free, near-adiabatic expansion, with a centre-of-mass
moving on a ballistic trajectory in the cluster potential.
As it interacts with the intracluster medium, this gas
can be shock heated to high temperatures. We note that in a
cluster the mean particle densities are very low, implying long
cooling times and a large fraction of $M_{vis}$ could therefore be
in the form of hot, diffuse gas. In consequence, phenomena such
as the Sunyaev-Zel'dovich (SZ) effect, which are contingent on
the existence of {\it tenuous\/} hot gas, are expected to be
present in our theory. Because we are ascribing a substantial
fraction of the observed X-ray emission to cloud collisions,
it is clear that the expected magnitude of the SZ effect
is diminished relative to the standard theory of cluster
X-ray emission. However, accurate measurements would
be necessary to distinguish between our theory and the
standard model, whereas the SZ effect has only recently been
convincingly detected at all (e.g. Rephaeli 1995). A referee
has brought to my attention the point that the spatial distribution
of SZ decrement can be used to test the proposed model. The
theory predicts that the diffuse, hot gas should lie mostly
within a radius of order $r_c$ of the cluster centre --- i.e.
more compact than is conventionally assumed. Existing images (e.g.
Carlstrom et al 1999) appear to be consistent with the proposed
model, in that they indicate characteristic radii comparable to
$r_c$ for the underlying gas distribution. We note that if the
theory presented here is correct then the SZ effect is unlikely to
prove useful as a technique for measuring the distances to clusters.

Although our theory predicts a similar mean intensity profile (eq. 4)
to that of the conventional model, the instantaneous distribution
consists of a large number of point-like sources, and the most
fundamental test of the theory would be to attempt to resolve the
emission from a cluster into its component sources. Each of these
sources should be transient, with a characteristic time-scale
$t_0\simeq r/2\sigma$, and for $r\sim1$~AU this is
$t_0\sim1/\sigma_3$~days. The mean luminosity is
$L_0\sim6.7\times10^{39}\,\sigma_3^3\;\ergs$ (this estimate
assumes a virial temperature of order 10~K for the clouds, cf.
Wardle \& Walker 1999); in turn this implies a total number
$\sim4.7\times10^4\sigma_3^{5/2}$ of sources contributing to the
cluster luminosity. In the case of the Virgo cluster, the
nearest rich cluster of galaxies ($\sigma\simeq650\kms$,
$D\simeq15$~Mpc), we deduce: a mean flux
of roughly $7\times10^{-14}\;{\rm erg\,cm^{-2}\,s^{-1}}$;
$t_0\sim1.5$~days; a total number of order 16,000 sources; and a
peak source density (in the cluster core) of $17\;{\rm arcmin^{-2}}$.
These estimates should be interpreted as order-of-magnitude estimates
only; nevertheless they indicate that the X-ray satellite Chandra should
easily detect individual transients within the Virgo cluster, even
in relatively short observations of an hour or so. By virtue of
Chandra's high resolution imaging, source confusion should not
be a problem even in the core of the cluster. The ROSAT
satellite was less sensitive than Chandra, and had much poorer
angular resolution, but even the ROSAT All Sky Survey (RASS) should
have revealed the brightest of the ongoing collisions at the
periphery of the Virgo cluster, where source confusion is
expected to be less of a problem than in the core. Inspection
of the publicly available RASS image of Virgo (reproduced
in figure 3) suggests that this is indeed the case, as the
outer regions of the cluster appear to possess a great deal of
compact substructure. These data are discussed in
B\"ohringer et al (1994).

What are the implications of non-detection of the predicted
population of transient sources in Virgo? Chandra's resolution
and sensitivity are such that the predicted sources should
be detectable, and not confused with each other, even if their
fluxes are an order of magnitude lower than the predicted value.
This is a sufficiently wide margin for error (in the modelling)
that observations with Chandra should be definitive: if Chandra
does not detect these sources, then clouds of the type we have
discussed make only a small contribution to the dark matter
in the Virgo cluster. As the collision rate, and hence the
expected number of detectable sources, is proportional to the
square of the number of clouds per unit volume, if no collisions
are observed where $10^4$ are expected, then the putative clouds
comprise $\lta1\,$\% of all the matter in Virgo.
Entities as large as clusters are widely regarded as
representative samples of the Universe as a whole, so
in turn this can be taken as a limit on the contribution
of cold clouds to the total matter density of the Universe. 

In principle, non-detection by Chandra admits another possible
interpretation: collisions which are so brief that less than five
photons are collected by Chandra from each event; this circumstance
would require cloud masses $M\lta7\times10^{-8}\msun$. This, however,
is not a self-consistent model: the condition
$\Sigma=134\;{\rm g\,cm^{-2}}$, from the Tully-Fisher relation
(\S2; W99), with the simultaneous requirement that the cloud
temperature be greater than the temperature of the microwave
background (3K), fixes a lower limit on the cloud mass of
$10^{-5}\msun$. Neither of these requirements can be relaxed
without compromising the model, so Chandra will provide a
strong test of the theory. We also note that the thermodynamic
(temperature) requirement alone demands very small radii,
$r\lta4\times10^{10}\;{\rm cm}$, for $M\lta7\times10^{-8}\msun$,
making it difficult to explain the extreme scattering events
(Walker \& Wardle 1998) if low-mass clouds are invoked.

An interesting qualitative point is that the X-ray spectra of
galaxy clusters typically exhibit iron abundances of order 0.3 (in
solar units: Mushotzky \& Loewenstein 1997). If a large fraction
of the X-ray emission does indeed arise from cloud-cloud
collisions, then these clouds presumably contain iron and
other heavy elements. (This conclusion can also be tested by
Chandra observations.) The simplest interpretation of this
point is that it indicates primordial non-zero heavy element
abundances. This is highly unconventional, but no more
so than the idea that all of the dark matter might be in the form of
cold gas clouds. Indeed, as emphasised by Walker \& Wardle (1999),
the two ideas are linked: if all of the dark matter is baryonic,
then consistency of the Big Bang nucleosynthesis calculations
with the observed {\it light\/} element abundances requires that
the Universe was inhomogeneous at the epoch of cosmic nucleosynthesis,
and in turn this admits the possibility of primordial {\it heavy\/}
element nucleosynthesis. This logic led Walker \& Wardle
(1999) to propose that the genesis of the (proto-)clouds
involved a phase transition in the very early Universe (i.e. prior to
the epoch of cosmic nucleosynthesis) -- cf. Hogan (1978).
The abundant iron in cluster X-ray spectra underlines this
interpretation, thereby connecting the cold-cloud model
directly to the physics of elementary particles. 

\section{Conclusions}
We have shown that if the dark matter is entirely composed of
cold gas clouds, then a substantial fraction of the observed X-ray
emission from clusters should be due to physical collisions between
these clouds. This possibility appears to be consistent with
existing data. Indeed the form of the observed cluster
luminosity-temperature correlation, and the measurement of high
EUV luminosities for some clusters, both suggest that this process
may well be occurring. If so then high-resolution
images of the Virgo cluster should reveal a large number of point-like,
transient X-ray sources contributing to the emission. Conversely, if
these sources are not seen, then cold clouds of $M\gta10^{-7}\;\msun$
cannot contribute more than about 1\% of the dark matter, either
in clusters or, by extension, in the Universe as a whole.

\section*{Acknowledgements}
I thank Mark Wardle for providing numerical polytropic density profiles,
and for several useful discussions. Andy Fabian, Ron Ekers
and Haida Liang contributed helpful advice on clusters.

\section*{References}

\reference  Arnaud~M., Evrard~A.E. 1999, MNRAS, 305, 631
\reference  Bloemen~H. 1989, ARAA,  27, 469
\reference  B\"ohringer~H., Briel~U.G., Schwarz~R.A., Voges~W.,
          Hartner~G., Tr\"umper~J. 1994 Nature 368, 828
\reference  Carlstrom~J.E., Joy~M.K., Grego~L., Holder~G.P., Holzapfel~W.L.,
          Mohr~J.J., Patel~S., Reese~E.D. 1999 ``Particle physics and
          the universe'' eds L.~Bergstrom, P.~Carlson, C.~Fransson
          (In press, astro-ph/9905255)
\reference  Clarke~C.J., Pringle~J.E. 1997, MNRAS, 288, 674
\reference  de Paolis F., Ingrosso~G., Jetzer~Ph., Roncadelli~M. 1995,
             Phys. Rev. Lett., 74, 14
\reference  Dixon~D.D., Hartmann D.H., Kolaczyk~E.D., Samimi~J.,
          Diehl~R., Kanbach~G.,\\ Mayer-Hasselwander~H.,
          Strong~A.W. 1998, New Ast., 3, 539
\reference  Draine~B.T. 1998 ApJL 509, L41
\reference  Fabian~A.C 1994, ARAA, 32, 277
\reference  Fabian~A.C., Crawford~C.S., Edge~A.C., Mushotzky~R.F. 1994,
          MNRAS, 267, 779
\reference  Fiedler R.L., Dennison~B., Johnston K.J., Hewish A. 1987,
          Nature, 326, 675
\reference  Gerhard~O., Silk~J. 1996, ApJ, 472, 34
\reference  Girardi~M., Fadda~D., Giuricin~G., Mardirossian~F.,
          Mezzetti~M., Biviano~A. 1996, ApJ, 457, 61 
\reference  Henriksen~R.N., Widrow~L.M. 1995 ApJ 441, 70
\reference  Hogan~C.J. 1978 MNRAS 185, 889
\reference  Jones~C., Forman~W. 1984, ApJ, 276, 38
\reference  Kaastra~J.S., Lieu~R., Mittaz~J.P.D., Bleeker~J.A.M.,
          Mewe~R., Colafrancesco~S.,\\Lockman~F.J. 1999, ApJL, 519, L119
\reference  Kalberla~P.M.W., Shchekinov~Yu.A., Dettmar~R.J. 1999, A\&A, 350, L9
\reference  Lieu R., Bonamente~M., Mittaz~J.P.D. 1999, ApJ, 517, L91
\reference  Markevitch M. 1998, ApJ, 504, 27
\reference  Mittaz~J.P.D., Lieu R., Lockman~F.J 1998, ApJ, 498, L17
\reference  Moore~B., Ghigna~S., Governato~F., Lake~G., Quinn~T, Stadel~J.,
          Tozzi~P. 1999 ApJL 524, L19
\reference  Mushotzky~R.F., Loewenstein~M. 1997, ApJ, 481, L63
\reference  Peebles~P.J.E. 1993 ``Principles of Physical Cosmology''
              (Princeton Univ. Press: Princeton)
\reference  Pfenniger D., Combes F., Martinet~L. 1994, A\&A, 285, 79
\reference  Rafikov~R.R., Draine~B.T. 2000 ApJ (submitted) astro-ph/0006320
\reference  Reichart~D.E., Castander~F.J., Nichol~R.C. 1999, ApJ, 516, 1
\reference  Rephaeli~Y. 1995, ARAA, 33, 541
\reference  Sarazin C.L. 1988 ``X-ray emissions from clusters of galaxies''
             (CUP: Cambridge)
\reference  Sarazin~C.L., Lieu~R. 1998, ApJ, 494, L177 
\reference  Shull~J.M., McKee~C.F. 1979, ApJ, 227, 131
\reference  Walker~M. 1999, MNRAS 308, 551 (W99)
\reference  Walker M., Wardle M. 1998, ApJ, 498, L125
\reference  Walker M., Wardle M. 1999, PASA, 16 (3), 262
\reference  Wardle M., Walker M. 1999, ApJL, 527, L109
\reference  Wu X.P., Xue Y.J., Fang~L.Z. 1999, ApJ, 524, 22

\end{document}